\pdfoutput=1
\RequirePackage{ifpdf}
\ifpdf 
\documentclass[pdftex]{sigma}
\else
\documentclass{sigma}
\fi

\numberwithin{equation}{section}

\def\na{\nabla}
\def\nab#1{{\buildrel #1\over \na}}
\def\sFrac[#1/#2]{\hbox{$\frac{#1}{#2}$}}
\def\Frac[#1/#2]{\frac{#1}{#2}}

\def\calR{{\cal R}}
\def\calH{{\cal H}}
\def\calC{{\cal C}}

\def\L{\hbox{\bf L}}

\def\R{{\mathbb R}}

\def\then{\Rightarrow}

\def\Ga{\Gamma}
\def\al{\alpha}
\def\be{\beta}
\def\de{\delta}
\def\si{\sigma}

\def\ep{\epsilon}
\def\vp{\varphi}
\def\te{\theta}
\def\Om{\Omega}
\def\la{\lambda}
\def\ga{\gamma}
\def\boldrho{\boldsymbol{\rho}}
\def\La{\Lambda}
\def\Si{\Sigma}

\begin{document}


\renewcommand{\thefootnote}{$\star$}

\newcommand{\arXivNumber}{1509.08008}

\renewcommand{\PaperNumber}{006}

\FirstPageHeading

\ShortArticleName{Extended Cosmologies}

\ArticleName{Extended Cosmologies\footnote{This paper is a~contribution to the Special Issue
on Analytical Mechanics and Dif\/ferential Geometry in honour of Sergio Benenti.
The full collection is available at \href{http://www.emis.de/journals/SIGMA/Benenti.html}{http://www.emis.de/journals/SIGMA/Benenti.html}}}

\Author{Salvatore {CAPOZZIELLO}~$^{\dag^1 \dag^2 \dag^3}$, Mariafelicia F.~{DE LAURENTIS}~$^{\dag^3 \dag^4 \dag^5 \dag^6}$,\\
 Lorenzo {FATIBENE}~$^{\dag^7 \dag^8}$, Marco {FERRARIS}~$^{\dag^7}$ and Simon {GARRUTO}~$^{\dag^7 \dag^8}$}

\AuthorNameForHeading{S.~Capozziello, M.F.~De Laurentis, L.~Fatibene, M.~Ferraris and S.~Garruto}

\Address{$^{\dag^1}$ Dipartimento di Fisica, University of Napoli ``Federico II'', Italy}
\EmailDD{\href{mailto:capozzie@na.infn.it}{capozzie@na.infn.it}}

\Address{$^{\dag^2}$~INFN Sezione  Napoli -- Iniziativa Specifica QGSKY, Italy}
\Address{$^{\dag^3}$~Gran Sasso Science Institute (INFN), L'Aquila, Italy}
\EmailDD{\href{mailto:mariafelicia.delaurentis@gmail.com}{mariafelicia.delaurentis@gmail.com}}
\Address{$^{\dag^4}$~Tomsk State Pedagogical University, Russia}
\Address{$^{\dag^5}$~Laboratory of Theoretical Cosmology, Tomsk State University of Control Systems \\
\hphantom{$^{\dag^5}$}~and Radioelectronics (TUSUR), Russia}
\Address{$^{\dag^6}$~Institut f\"ur Theoretische Physik, Goethe-Universit\"at, Max-von-Laue-Str.~1,\\
\hphantom{$^{\dag^6}$}~60438 Frankfurt, Germany}
\Address{$^{\dag^7}$~Dipartimento di Matematica, University of Torino, Italy}
\EmailDD{\href{mailto:lorenzo.fatibene@unito.it}{lorenzo.fatibene@unito.it}, \href{mailto:marco.ferraris@unito.it}{marco.ferraris@unito.it}, \href{mailto:simon.garruto@unito.it}{simon.garruto@unito.it}}
\Address{$^{\dag^8}$~INFN Sezione  Torino -- Iniziativa Specifica QGSKY, Italy}

\ArticleDates{Received September 29, 2015, in f\/inal form January 16, 2016; Published online January 20, 2016}

\Abstract{We shall discuss cosmological models in extended theories of gravitation.
We shall def\/ine a surface, called the {\it model surface}, in the space of observable parameters which characterises families of theories.
We also show how this surface can be used to compare with observations.
The model surface can potentially be used  to falsify whole families of models
instead reasoning on a single model basis as it is usually done by best f\/it arguments with observations.}


\Classification{83F05; 83D05; 37N20}

\renewcommand{\thefootnote}{\arabic{footnote}}
\setcounter{footnote}{0}

\section{Introduction}

The aim of this paper is to provide a detailed framework for interpreting extended cosmologies in the family of
Palatini $f(\calR)$-models.
The interpretation of extended theories is trickier than usually recognised in the literature,
since the dynamics is not suf\/f\/icient to constrain all details of the interpretation
or the relation between the cosmographic parameters and the corresponding model quantities.
All {interpretation issues}  are generated by the fact that in Palatini $f(\calR)$-theories one has two metrics
which play a natural role, the original metric $g$ and a conformal metric $\tilde g$ which naturally appears  when solving the model.

From the geometric point of view, while in standard general relativity (GR) the geometry of spacetime is described by a Lorentzian structure, in $f(\calR)$-theories spacetime is described by a Weyl geometry, i.e., by a conformal structure~$[g]$ together with a compatible connection {$\tilde\Ga$} which, in view of f\/ield equations, turns out to be the Levi-Civita connection of the conformal metric $\tilde g{=\vp\cdot g}$.
The scalar f\/ield $\vp$ is called the {\it conformal factor} and it parametrises conformal transformations.
Weyl geometries have been considered in the past in gravitational theories; see, for example,~\cite{R2a,Perlick,R2c,R2b,R2d, Weyl}.

Unlike in standard GR, where one has only one metric which does everything, in Palatini $f(\calR)$-theories
one has to declare which metric is used to do what; see, for example,~\cite{Ba, Bb,Bc}.
For example, the dynamics of f\/ield equations does not determine the free fall.
If one couples to a~matter f\/ield of course one can compute the characteristics of matter f\/ield equations and argue that characteristic curves describe free fall.
In particular, if one couples a Klein--Gordon scalar f\/ield $\phi$ to the metric~$g$ then characteristics are determined as the geodesics of~$g$.
Unlike lightlike geodesics, the timelike geodesics of~$g$ are {not} geodesics of~$\tilde g$.
Accordingly, it seems that the free fall is determined by the variational principle assumed for gravity and matter.

However, one can show that a conformal transformation can be extended to  act on the matter f\/ield $\phi$ as well and def\/ine a new matter f\/ield
$\tilde \phi = \vp^\al \cdot \phi$ (with the parameter $\al\in\R$ to be determined) so that the new f\/ield~$\tilde \phi$ obeys new (dif\/ferent) f\/ield equations (since conformal transformations are not symmetries of the matter f\/ield equations). The exponent~$\al$ in the {action of the} conformal transformation on the Klein--Gordon f\/ield can be chosen so that the characteristics of the new f\/ield~$\tilde \phi$ are geodesics of the conformal metric~$\tilde g$; see~\cite{Corfu}.
Something similar can be done for Dirac f\/ields.

Hence  arguing that the variational principle determines the free fall is correct, {\it provided} one assumes  what is observed as a matter f\/ield.
Depending on the choice of~$\phi$ or $\tilde \phi$ as the observed f\/ield, then free fall is determined to be described by~$g$ or~$\tilde g$, respectively.
In this paper we shall assume that free fall is given by~$\tilde g$.

Other examples are the issues of whether the
dynamics determines the metric which is used for def\/ining spatial lengths, minimal coupling to matter, standard clocks.
For example, in standard GR it is natural assuming that physical lengths are associated to the geometric length measured with respect to the only metric one has around, namely~$g$.
Having more that one metric in the model makes the choice much more problematic.

Let us consider two parallel mirrors at a constant distance and a light ray bouncing between them, which is called a {\it gravitational clock}.
If one considers an {\it atomic clock} (which is in fact built with some quantum harmonic oscillator)
 and compares its rate to the rate of a gravitational clock, the assumption of standard GR is pretty neat: the two clocks (at the same point) are synchronised forever.

It is quite obvious that the discrepancy between the two clocks, if any, cannot be too big other\-wise we shall see some ef\/fect at the solar system scales. However, there is no strong constraint on what happens at bigger scales.
One should not {\it assume} that there is no discrepancy. On the contrary, the correct procedure would be to assume there is, derive observational consequences, and constrain the ef\/fects by observations.

\looseness=-1
We shall hereafter follow this second guideline. In this paper we shall assume that lengths are related to $g$, not to~$\tilde g$.
Accordingly, we shall assume that matter is coupled to~$g$, and not to~$\tilde \Ga$.

These choices are also motivated by the Ehlers--Pirani--Schild (EPS) framework for gravitational theories; see~\cite{EPS}.
In the early seventies EPS proposed a framework able to derive the geometry of spacetime based on basic properties of light rays and (free falling) particles. In this framework one does not assume a Lorentzian structure on spacetime,
rather a Lorentzian metric (actually a Lorentzian conformal structure) on spacetime can be def\/ined
due to the properties assumed for light rays.
As a consequence, in the EPS framework there is a~tight link between the geometric structure of spacetime and observations.
One could say that the EPS framework instructs an observer to establish observational protocols and to def\/ine distances and time lapses.
Once this is done EPS considered the properties of particles and f\/ind they are described by a connection~$\tilde \Ga$
(actually by a projective structure on the spacetime) which describes, by construction, the free fall.

The two structures $([g], \tilde \Ga)$ on spacetime are not arbitrary though. Since one knows that lightlike directions are
an upper bound for {the speed of particles}, one can show that the connection~$\tilde \Ga$ must be in the form
\begin{gather}
\tilde \Ga^\al_{\be\mu}= \{g\}^\al_{\be\mu} + \sFrac[1/2]\big(g^{\al\ep} g_{\be\mu} - 2 \de^\al_{(\be} \de ^\ep_{\mu)}\big) A_\ep
\label{EPSComp}
\end{gather}
for some covector $A_\ep$.
This is quite similar to the original proposal by Weyl for a unif\/ied theory of gravity and electromagnetism (then proven wrong); see~\cite{Weyl}.
However, we are not interpreting here~$A$ as an electromagnetic potential
and for us the covector~$A_\ep$ will be related to the possible shift in synchronisation between atomic and gravitational clocks.

On the basis of the EPS framework we shall consider gravitational theories for the f\/ields $(g, \tilde \Ga)$
to be {\it extended theories of gravitation} {(ETG)} when their f\/ield equations enforce as a consequence the relation~(\ref{EPSComp});
see~\cite{Cap1, Cap2}.
We shall call  such theories {\it integrable} when f\/ield equations also imply that $A$ is {an exact}  $1$-form and as a consequence~$\tilde \Ga$ is a metric connection; see \cite{Extended1,Extended2}.

In principle one can consider any invariant of the curvature to def\/ine a dynamics.
We shall hereafter restrict to Palatini $f(\calR)$-theories in which the dynamics is given in terms of a (regular) function
of the scalar curvature~$\calR$.
One can show that  Palatini $f(\calR)$-gravity is an integrable extended theory of gravitation (iETG).
In view of the EPS framework we shall interpret $\tilde \Ga$ as def\/ining the free fall, while $g$ is related to distances and time lapses.

Standard GR is then a particular case of Palatini $f(\calR)$-theory. However, the standard Hilbert--Einstein dynamics imposes
that the conformal factor {$\vp$}  is exactly and everywhere equal to~1.
In this special case  the two metrics $g$ and $\tilde g$ do coincide and all issues related to choosing which metric is used to do what are overcome.

It is particularly worth noticing that the EPS project was originally started to show that standard GR follows from assumptions
and it almost reached the goal. Instead of f\/inding a~Lorentzian structure on the spacetime as assumed in standard GR,
EPS f\/ind something less (just a conformal structure of Lorentzian metrics) and something more (a~projective structure of connections which {is}  not uniquely determined as the Levi-Civita connection of the metric structure).

EPS asked in their analysis what one needs to  add to the framework in order to force the structure to be Lorentzian.
They showed that in fact if one requires that atomic and gravitational clocks have the same rate then standard GR follows.

In view of this, Palatini $f(\calR)$-theories can be seen to generalise standard GR in two respects.
In the f\/irst place, the modif\/ied dynamics induces (ef\/fective) dark sources which can be used to model dark matter and dark energy.
In the second place, the conformal factor does account for the possibility that gravitational clocks shift with respect to atomic clocks, the ef\/fect being measured by the f\/ield~$\vp$, i.e., the conformal factor; see \cite{Cap3, Polistina}.

These two ef\/fects (dark sources and time shifts) are quite independently assumed in the theory and in our opinion
one should not  {\it assume} they are not there.
Instead, they  should be analysed by considering them in a model, {one should} make predictions and test the results, possibly ruling them out.

The material is organised as follows.
In Section~\ref{section2} we shall review Palatini $f(\calR)$-theories.
The conservation of stress tensors and their relation with conformal transformations is discussed in Appendix~\ref{appendixA}.
In Section~\ref{section3} we shall discuss cosmological models based on  Palatini $f(\calR)$-theories.
In Section~\ref{section4} we shall consider the example $f(\calR)=\calR -\Frac[\ep/2]\calR^2$, the solutions of which are brief\/ly discussed in the Appendix~\ref{appendixB}.

\section[Palatini $f(\calR)$-theories of gravitation]{Palatini $\boldsymbol{f(\calR)}$-theories of gravitation}\label{section2}

In view of the EPS formalism one should consider a Palatini formulation.
For the sake of generality, one can consider a general dynamics, not restricting {\it a priori} to standard~GR.
For the sake of simplicity let us consider a Palatini $f(\calR)$-dynamics, and we shall assume
for it the interpretation described above.

Thus our theory is described by two f\/ields $(g_{\mu\nu}, \tilde \Ga^\al_{\be\mu})$
a metric $g$ which is associated with rulers and clocks and a (torsionless) connection~$\tilde \Ga$
which is associated to free fall of test particles.

Let us def\/ine the scalar curvature
$\calR= g^{\mu\nu} \tilde R_{\mu\nu}(\tilde \Ga)$
and the Lagrangian
\begin{gather*}
\L=  f(\calR)  \sqrt{g} d\si + \L_{\rm mat},
\end{gather*}
where $d\si$ denotes the standard local pointwise   basis of $4$-forms on a $4$-dimensional spacetime~$M$
induced by coordinates.
The  f\/ield equations turn out to be
\begin{gather*}
 f'(\calR) \tilde R_{(\mu\nu)} -\sFrac[1/2] f(\calR)g_{\mu\nu}= T_{\mu\nu},\\
 \tilde \na_\al\big(\sqrt{g} f'(\calR) g^{\mu\nu}\big)=0,
\end{gather*}
where the stress tensor~$T_{\mu\nu}$  is def\/ined by variation of matter Lagrangian
with respect to the metric~$g$. We also assumed that the matter does not couple to the connection~$\tilde \Ga$
but only to the metric~$g$ {and we understand the matter f\/ield equations}.

The second equation does determine the connection $\tilde \Ga$ as the Levi-Civita connection of a~conformal metric
\begin{gather*}
\tilde g_{\mu\nu}:= \vp\cdot g_{\mu\nu},
\qquad \vp=\Frac[ f'(\calR)/ \vp_0]
\qquad\then \qquad
\tilde \Ga= \{\tilde g\}
\end{gather*}
for some constant $\vp_0$ chosen so that one has $\vp(x_0)=1$ here and now.
{Let us assume} that the conformal factor $\vp$ is positive {and} the metric $\tilde g$ is also Lorentzian.
Using this information in the f\/irst equation, the Ricci tensor $\tilde R_{\mu\nu}$ becomes the (symmetric) Ricci tensor of the conformal metric~$\tilde g$, i.e., one obtains
\begin{gather*}
f'(\calR) \tilde R_{\mu\nu} -\sFrac[1/2] f(\calR)g_{\mu\nu}= T_{\mu\nu}.
\end{gather*}

In order to solve this equation let us consider its trace by $g^{\mu\nu}$ f\/irst
\begin{gather}
f'(\calR)  \calR - 2 f(\calR)=  T,
\label{MasterEquation}
\end{gather}
which is called the {\it master equation}.
It is an algebraic (i.e., not dif\/ferential) equation in $\calR$ and $T= g^{\mu\nu} T_{\mu\nu}$;
except some degenerate cases, for a generic (analytic) $f(\calR)$ one can solve it for $\calR$ obtaining
\begin{gather*}
\calR= \calR(T).
\end{gather*}
Accordingly, one can express the conformal factor as a function of the trace~$T$, i.e.,
\begin{gather*}
\vp(T)= \Frac[f'(\calR(T))/\vp_0].
\end{gather*}

Using again this information to write the f\/irst f\/ield equation in terms
of the new conformal metric $\tilde g$ one has
\begin{gather*}
\tilde R_{\mu\nu} -\sFrac[1/2] \tilde R \tilde g_{\mu\nu}=
 \tilde T_{\mu\nu}
:=  \sFrac[1/\vp] T_{\mu\nu}  +  \sFrac[1/2\vp] f(\calR)g_{\mu\nu} - \sFrac[1/2]\calR  g_{\mu\nu},
\end{gather*}
where $\tilde R=\vp^{-1} \calR$ is the scalar curvature of the conformal metric.
In other words, the conformal metric $\tilde g$ is determined by a Einstein equation though with a modif\/ied source stress tensor
\begin{gather}
\tilde T_{\mu\nu}:= \sFrac[1/\vp] T_{\mu\nu}  +    \sFrac[1/2\vp] f(\calR)g_{\mu\nu} - \sFrac[1/2]\calR  g_{\mu\nu},
\label{EffectiveT}
\end{gather}
which can in fact be considered as a function of $\tilde g$ and the matter f\/ields
(included the conformal factor~$\vp$).

As a minimal check one can show that if the matter energy-momentum stress tensor $T_{\mu\nu}$ is conserved (with respect to~$g$) then the ef\/fective stress tensor~$\tilde T_{\mu\nu}$ is automatically conserved as well, with respect to~$\tilde g$, though.
See Appendix~\ref{appendixA} for details.

\section[Palatini $f(\calR)$-cosmologies]{Palatini $\boldsymbol{f(\calR)}$-cosmologies}\label{section3}

We can use this theory of gravitation for producing cosmological models.
For, one has to state {\it cosmological principle}, i.e., restricting to spacetimes with an isometry group
which enforces spatial homogeneity and isotropy (or, equivalently, spatial isotropy with respect to any point).
As Benenti pointed out, the cosmological principle freezes dynamics since there is only one dynamics,
namely the one dictated by {\it Friedman--Robertson--Walker} (FRW) equations,
which is compatible with homogeneity and isotropy; see~\cite{Benenti}.

However, if, as in this case, a theory can be described in terms of more than one metric, one has to state for which metric isometries are considered in the cosmological principle.
Luckily, in Palatini $f(\calR)$-theories this is not an issue, {since the conformal factor $\vp(t)$ itself is expected to be a function of time only}, there are two metrics which are conformal to each other by a~conformal factor which depends on time only,
and if a metric $g$ is spatially homogeneous and isotropic then the conformal metric~$\tilde g$
is spatially homogeneous and isotropic as well, though in a particular and precise sense which we need to make explicit
for later convenience.

A metric $g$ obeys cosmological principle if and only if there exists a coordinate system $(t, r, \te, \phi)$ in which the metric is in the form
\begin{gather}
g=-  dt^2 +   a^2(t) \left(\Frac[dr^2/1-k r^2] + r^2 d\Om^2\right),
\qquad
d\Om^2:= d\te^2 + \sin^2(\te) d\phi^2
\label{FRWg}
\end{gather}
for some function of time $a(t)$ which is called the {\it scale factor}.
A metric in this form is called a~{\it FRW metric}.
One can (and we will) always rescale the radial coordinate so to have $a(t_0)=1$ at a time $t_0$ denoting today (which for convenience can be set to $t_0=0$).

If the conformal factor is a function of time only the conformal transformation sends a FRW metric into a FRW metric, though with another time and scale factor.
If~$g$ is in the form~(\ref{FRWg}) with radial coordinate rescaled to obtain~$a(t_0)=1$,
then the conformal metric $\tilde g= \vp(t) \cdot g$ is in the form
\begin{gather}
\tilde g= -   d{\tilde t}{}^2 +   \tilde a^2(\tilde t) \left(\Frac[d r^2/1- k  r^2] +  r^2 d\Om^2\right)
\qquad\then\qquad
\begin{cases}
\tilde a(\tilde t) = \sqrt{\vp(t)}  a(t),\\
d{\tilde t}= \sqrt{\vp(t)} d t,
\end{cases}
\label{ConformalMetricFRW}
\end{gather}
and one still has  $\tilde a(\tilde t_0)=1$ {if and only if the conformal factor~$\vp(t)$ is chosen so that $\vp(t_0)=1$}.

Most of what follows relies on having a metric in FRW form and, consequently, it applies to both~$g$ and~$\tilde g$.
We shall hereafter brief\/ly review what is routinely done for~$g$ in standard cosmology with the aim of then applying it to $\tilde g$.

Once one has a FRW metric $g_{\mu\nu}$, one can compute its Einstein tensor $G_{\mu\nu}$ and f\/ind it in the form of a~stress tensor of a perfect f\/luid $T_{\mu\nu}$, i.e., there exists a vector f\/ield~$u$
which is a timelike unit  vector for the metric~$g$ orthogonal to the submanifolds at $t={\rm const}$
and two functions~$p(t)$ (the {\it pressure})  and $\rho(t)$ (the {\it energy density}) such that
\begin{gather}
T_{\mu\nu}=  (\rho + p )u_\mu u_\nu + p g_{\mu\nu}
\qquad\then\qquad
T=-(\rho+ p) + 4 p= 3p-\rho.
\label{FluidT}
\end{gather}

This has not much to do with dynamics since the stress tensors for {a} perfect f\/luid are the only tensors  which are compatible with spatial homogeneity and isotropy. In some sense the cosmological principle freezes the dynamics and reduces it to a~purely kinematical fact:
any FRW metric obeys Einstein equations with a perfect f\/luid source for a suitable pressure and energy density functions.
The dynamics only specif\/ies the relation of density and pressure with the matter f\/ields.

Einstein equations are $10$ equations, though in this case only two {of them} are independent.
They are the so-called  {\it Friedman equations}
\begin{gather}
\left(\Frac[\dot { a}/  a]\right)^2 + \Frac[k c^2/ a^2]= \Frac[1/3] \rho, \qquad
\Frac[\ddot { a}/ a]= -\Frac[1/6] (\rho+3  p).
\label{FRWeqs}
\end{gather}

These are two equations for the three functions $a(t)$, $\rho(t)$, $p(t)$ which then cannot be determined
unless a further equation is provided.
In principle one should obtain that {extra} equation from the matter f\/ield equations.
In practice, it is customary in cosmology to describe matter by means of an (algebraic) equation of state~(EoS)
which constrains~$\rho$ and~$p$.
The most general EoS $\si(p, \rho)=0$ usually is chosen in the form $p=p(\rho)$ and often linearised
simply to~$p=w\rho$, where $w$ is called the {\it parameter of the EoS}, the value of which determines the kind of matter we are dealing with. Dif\/ferent cases will be analysed below.

Equations (\ref{FRWeqs}) are Lagrange equation for a suitable Lagrangian.
If we consider the matter described by an energy-density function~$\rho(a)$,
then we can def\/ine  the so-called {\it point-like Lagrangian}
\begin{gather}
L=\big( {-} a \dot a^2 + kc^2 a -\la \rho(a) a^3\big) dt,
\label{FRWLag}
\end{gather}
which is obtained by integrating on space the Hilbert Lagrangian for a FRW metric (and neglec\-ting a total time-derivative).

This Lagrangian is time-independent so that its ``total energy''
\begin{gather*}
\calH= -2 a \dot a^2+ a \dot a^2 - kc^2 a +\la \rho(a) a^3=  -\left( \Frac[\dot a^2/a^2] +\Frac[ kc^2/a^2] - \la \rho(a) \right)a^3
\end{gather*}
is conserved.
Notice that by imposing this quantity to be zero one obtains
\begin{gather*}
\Frac[\dot a^2/a^2] +\Frac[ kc^2/a^2] = \la \rho,
\end{gather*}
which agrees with the f\/irst FRW equation by setting $\la=\sFrac[1/3]$.
The equation of motion for the Lagrangian~(\ref{FRWLag}) is
\begin{gather*}
 2\ddot a=\Frac[1/3]  ( 2 \rho  +  \rho' a )a \qquad\then\qquad
  \Frac[\ddot a/a] =\Frac[1/6]  ( 2 \rho  +  \rho' a ),
\end{gather*}
which becomes the second FRW equation provided one has the pressure def\/ined to be
\begin{gather*}
-( \rho + 3p) =2 \rho  +  \rho' a
\qquad\then\qquad
 p=- \big( \rho  + \sFrac[1/3] \rho' a\big).
\end{gather*}
This condition is equivalent to the conservation of the energy-momentum stress tensor, namely $\na_\mu T^{\mu\nu}=0$
for a f\/luid tensor in the form~(\ref{FluidT}).

Thus we select the type of matter f\/luid by choosing the function $\rho(a)$ which in turn determines the matter  equation of state (EoS) $p=p(\rho)$.
The def\/inition of the pressure above is in fact a~form for the EoS parametrised by the scale factor~$a$
\begin{gather*}
\rho=\rho(a), \qquad
p=p(a):=  -\big( \rho(a)  + \sFrac[1/3] \rho'(a) a\big).
\end{gather*}

For example, {\it dust} is determined by setting $p=0$, i.e., $3\rho  + \rho' a=0$ from which
$\rho= \rho_0 a^{-3}$ follows.
More generally, assuming an EoS in the form $p=w\rho$ one has
\begin{gather*}
\rho = \rho_0 a^{-3(w+1)}.
\end{gather*}
{\it Radiation} is def\/ined to be $p=\sFrac[1/3]\rho$, thence $\rho= \rho_0 a^{-4}$.
{\it Cosmological constant} or ({\it quintessence}) is def\/ined to be $p=-\rho$, so that one has $\rho= \rho_0$; see~\cite{Cap3}.

One can also assume the so-called polytropic equation of state $p=\rho^\ga$ and obtain
\begin{gather*}
\rho(a) =\left( \Frac[3/(ca)^{3(\ga-1)} -3]\right) ^{\sFrac[1/\ga-1]}.
\end{gather*}

In view of the point-like Lagrangian, FRW cosmology is captured by f\/irst FRW equation which is a Weierstrass equation:
\begin{gather}
\Frac[\dot a^2/a^2] +\Frac[ k/a^2] - \sFrac[1/3]  \rho(a)=0
\qquad\then\qquad
\dot a^2 =  - k  + \sFrac[1/3] \rho(a) a^2=:\Phi(a).
\label{FRW1}
\end{gather}
This function $\Phi(a)$ is essentially the so-called {\it effective potential}. The quantity
\begin{gather*}
H= \Frac[\dot a/a]
\end{gather*}
is called the {\it Hubble parameter}. The value of the Hubble parameter today is denoted by~$H_0$.

One can observe that if $k=0$ in equation~(\ref{FRW1}) then $\rho_0^{\rm cr}=3 H_0^2$ which is called the {\it critical density} (today).
Hence if one has dif\/ferent matter species each with its own density $\rho_i=\rho_i^0 a^{n_i}$ and pressure~$p_i$, and its own EoS $p_i=p_i(\rho_i)$,
the equation~(\ref{FRW1}) specif\/ies to
\begin{gather*}
\left(\Frac[\dot a /a]\right)^2 = \Frac[1/3] \left( \rho -\Frac[3k/a^2]\right)= \Frac[1/3]  \left(\sum_{i=1}^k \rho_i +\rho_0\right)=\Frac[1/3] \boldrho,
\end{gather*}
where we introduced one more species, called the {\it curvature species}, with density
$\rho_k = -3ka^{-2}$.
The {\it total density} $\boldrho=\sum\limits_{i=0}^k\rho_i$  must be critical today.

Then one can  recast the FRW equation as
\begin{gather*}
\Frac[\dot a^2/a^2] = \sFrac[1/3]   \rho_0^{\rm cr}  \sum_i \Frac[\rho_i/\rho_0^{\rm cr}]= H_0  \sum_{i=0}^k \Om_i^0 a^{n_i},
\end{gather*}
where we set
\begin{gather*}
\Om_i^0= \Frac[\rho^0_i/\rho_0^{\rm cr}],
\qquad
\sum_{i=0}^k \Om_i^0=1,
\end{gather*}
which are called {\it abundances} (today) and are one-dimensional parameter for each species.
Since a~set of abundances are constrained by $\sum\Om^0_i=1$, they can be interpreted as percentages of each species in the total budget
corresponding to the critical density. A set of such percentages are called a~{\it cosmic pie}.
Then in this more general framework, the FRW equation reads as
\begin{gather*}
\dot a^2= a^2 H_0 \sum_i \Om_i^0 a^{n_i} =\Phi\big(a; H_0,  \Om_i^0\big)
\end{gather*}
and solutions are uniquely determined by initial condition $a(0)=1$ and today abundances $\Om_i^0 $.

\subsection{Einstein frame}

In Palatini $f(\calR)$-cosmologies, also the conformal metric~$\tilde g$ is a FRW metric and as such it obeys
 Friedman equations for some suitable pressure and density as well.
We already showed that the conformal metric can be recast in FRW form by a suitable rescaling.

Thus we have the metric in the from~(\ref{ConformalMetricFRW}).
Since it is a FRW metric, it obeys some sort of Friedman equations which means that the modif\/ied stress tensor
$\tilde T_{\mu\nu}$ must be in the form of a~perfect f\/luid stress tensor.

In fact the timelike unit vector $\tilde u$ for the metric $\tilde g$ is
\begin{gather*}
{\tilde u}^\mu \tilde g_{\mu\nu} {\tilde u}^\nu = -1,
\qquad
u^\mu=\sqrt{\vp} \tilde u^\mu
\quad\iff\quad
u^\mu g_{\mu\nu} u^\nu=-1.
\end{gather*}
The corresponding covector is
\begin{gather*}
\tilde u_\mu= \tilde g_{\mu\nu} \tilde u^\nu=  \vp  g_{\mu\nu} \tilde u^\nu= \sqrt{\vp} g_{\mu\nu} u^\nu= \sqrt{\vp} u_\mu
\end{gather*}
and the modif\/ied stress tensor {(\ref{EffectiveT})} is in the form
\begin{gather*}
\tilde T_{\mu\nu} =
\Frac[p+\rho /\vp^2] \tilde  u_\mu \tilde u_\nu + \Frac[1/4\vp^2] (  p+\rho  -   \vp\calR     )\tilde g_{\mu\nu},
\end{gather*}
where the master equation has been used to eliminate the dependence on $f(\calR)$.

Thus also $\tilde T_{\mu\nu}$ is in canonical form provided one def\/ines the ef\/fective pressure~$\tilde p$ and energy density~$\tilde \rho$ as
\begin{gather*}
\tilde p + \tilde \rho= \Frac[p+\rho /\vp^2]
\qquad\then\qquad
\tilde \rho= \Frac[3(p+\rho) /4\vp^2] + \Frac[\calR/4\vp],
\qquad
\tilde p=  \Frac[p+\rho/4\vp^2]  - \Frac[\calR/4\vp].
\end{gather*}

By replacing $\calR= \calR(T)= \calR(T)= \calR(3p-\rho)$ and $\vp= \vp_0^{-1}f'(\calR(T))=\vp_0^{-1}f'(\calR(3p-\rho))$
and using the EoS, one obtains the ef\/fective pressure and energy density as
functions of~$\rho$, i.e.,
\begin{gather*}
\tilde \rho= \tilde \rho(\rho), \qquad
\tilde p=  \tilde p(\rho),
\end{gather*}
which is an implicit form of the EoS for the ef\/fective matter (parametrised {by} the matter energy density~$\rho$).

\section{An example}\label{section4}

Let us consider the simple case (see~\cite{Olmo})
\begin{gather*}
f(R)= \calR - \sFrac[\ep/2] \calR^2,
\qquad \ep\ge 0.
\end{gather*}

\begin{figure}[t] 
   \centering
   \includegraphics[width=5cm]{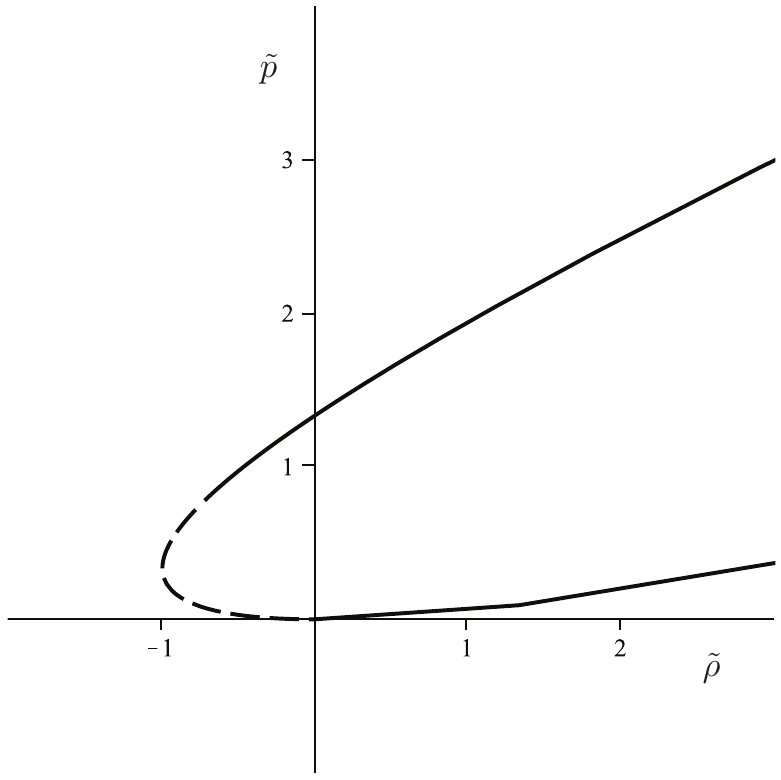}
   \qquad
      \includegraphics[width=6cm]{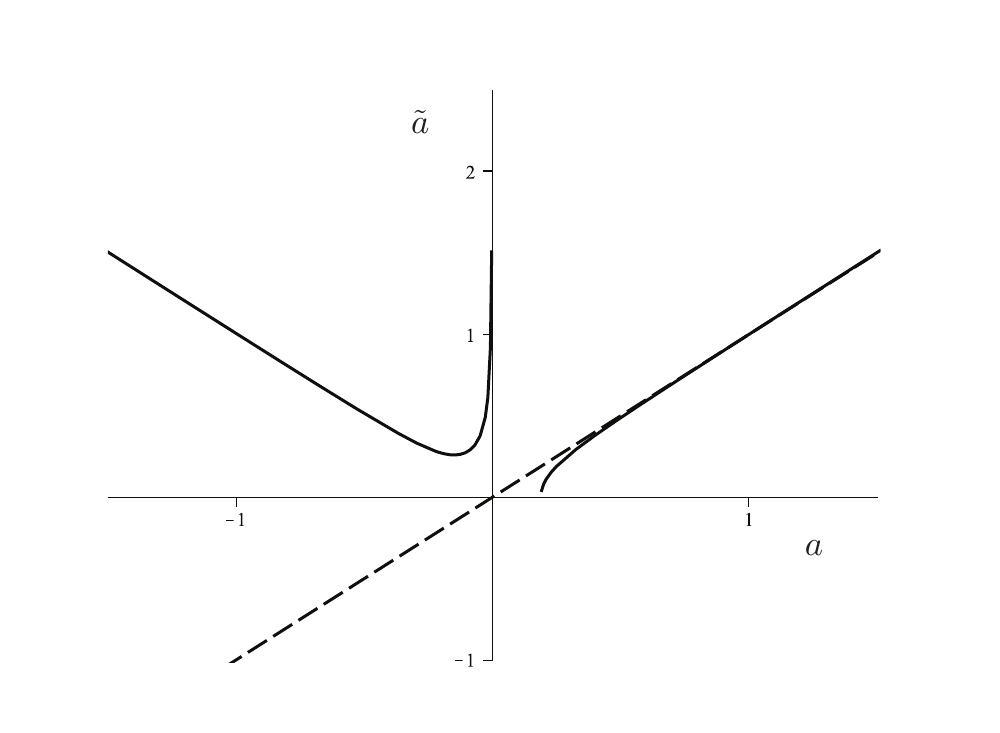}
   \caption{a) Ef\/fective EoS $ \tilde p(\tilde \rho)$. (The dash line corresponds to a~negative energy density~$\rho$ for visible matter, which may be considered unphysical.)
b)~Scale factors~$\tilde a(a)$. Dash line is standard GR.}   \label{fig1}
\end{figure}

The stress tensor $T_{\mu\nu}$ is assumed to be a f\/luid stress tensor.
Visible matter will be described by this tensor and, for the sake of simplicity, we will assume visible matter to be dust.
The master equation reads as
\begin{gather*}
\calR - \ep\calR^2 -2\calR +\ep \calR^2= -\calR=- \rho
\qquad\then\qquad
\calR=\rho
\end{gather*}
and the conformal factor is
\begin{gather*}
\vp=\Frac[1-\ep\calR/ \vp_0]= \Frac[1- {\ep} \rho/ 1- {\ep} \rho_0],
\qquad
\rho=\rho_0 a^{-3}.
\end{gather*}
Let us stress that there are two allowed regions for the scale factor, one for $a\ge  a_m=(\ep\rho_0)^{1/3}>0$ and one for negative values of~$a$.
The ef\/fective pressure and energy density are
\begin{gather*}
\tilde \rho= \Frac[3+\vp/4\vp^2]  \rho
=  \Frac[1-\ep \rho_0/(1 -
{\ep}\rho)^2] \big(1 - \sFrac[3/4]\ep\rho_0 -\sFrac[1/4] \ep\rho\big)\rho,
\\
\tilde p=  \Frac[1-\vp/4\vp^2]   \rho=
\Frac[\ep /4 ]\Frac[1-\ep \rho_0  /(1-\ep \rho)^2]  (\rho-\rho_0)\rho.
\end{gather*}
This corresponds to an ef\/fective EoS; see Fig.~\ref{fig1}a.

The relation between the scale factors is (see Fig.~\ref{fig1}b)
\begin{gather*}
\tilde a (a) = \sqrt{ \Frac[{1-   \ep \rho}/{ 1-   \ep \rho_0}] } a.
\end{gather*}

At this point we can expand $\tilde \rho$ and $\tilde p$ as functions of $a$ and check that
\begin{gather}
\tilde p = -\left(\tilde \rho +\sFrac[1/3]\sFrac[d\tilde \rho/d \tilde a] \tilde a\right),
\qquad
\Frac[d\tilde \rho/d \tilde a]= \Frac[d\tilde \rho/d \rho] \Frac[d \rho/d  a] \Frac[d a/d \tilde a]
= -3\rho_0 a^{-4}\Frac[d\tilde \rho/d \rho] \left(\Frac[d\tilde a/d  a]\right)^{-1}.
\label{WeierstrassCondition}
\end{gather}
The condition (\ref{WeierstrassCondition}) is in fact generally equivalent to conservation laws for the ef\/fective energy-momentum stress tensor $\tilde T_{\mu\nu}$, which is general follows from conservation of the original  energy-momentum stress tensor $T_{\mu\nu}$ (see Appendix~\ref{appendixA}), which in turn follows from EoS imposed for matter, or, equivalently, from covariance of the matter Lagrangian.

Then one decides what kind of ef\/fective matter is needed in the model. This is not something decided within cosmology, it can be a choice of ours or an input from other physics (e.g., constraints from baryogenesis or structures formation).

For example, let us assume we want a model with cosmological constant, dust and radiation. This  corresponds to a total density in the form
\begin{gather*}
\hat \rho(\tilde a)= \tilde\rho{}_\La^0 + \Frac[\tilde\rho{}^0_d/\tilde a^3] + \Frac[\tilde\rho{}^0_r/\tilde a^4]
\end{gather*}
with the parameters $(\tilde\rho{}_\La^0 , \tilde\rho{}^0_d, \tilde\rho{}^0_r)$ to be determined so that $\hat\rho$
best approximates $\tilde \rho$.

Best f\/it can be obtained simply by imposing that the value, f\/irst and second derivatives of $\tilde \rho$ and $\hat \rho$
coincides at $\tilde a=1$, i.e., today.
The best approximation takes the form of
\begin{gather}
\tilde\rho{}_\La^0 = \tilde\rho{}_\La^0 (\ep,\rho_0),
\qquad
\tilde\rho{}^0_d= \tilde\rho{}^0_d(\ep,\rho_0),
 \qquad
 \tilde\rho{}^0_r= \tilde\rho{}^0_r(\ep,\rho_0).
\label{Embed}
\end{gather}

This functions are quite complicated functions, also for {a} simple model as the one we are considering here.
However, one can have an idea of their shape (and the ef\/fective physics they describe) by expanding them for small
$\ep$ (i.e., {\it small} modif\/ication of standard GR).
In this case we have:
\begin{gather*}
\tilde\rho{}_\La^0 \simeq  \Frac[  \rho_0^2/8] \ep -\Frac[\rho_0^3/4]\ep^2,
\qquad
\tilde\rho{}^0_d\simeq  \rho_0
-\Frac[5 \rho_0^2/4]\ep
+\Frac[11 \rho_0^3/8]\ep^2,
 \qquad
 \tilde\rho{}^0_r\simeq
 \Frac[9  \rho_0^2/8]\ep
 -\Frac[9 \rho_0^3/8]\ep^2.
\end{gather*}
Let us remark that those reduce to visible matter (dust only) for $\ep=0$ and that all energy densities are positive for a small positive~$\ep$.

This means that Palatini $f(\calR)$-cosmology with $f(\calR)=\calR - \sFrac[\ep/2]\calR^2$
and purely dust visible matter is well approximated {\it today}
by standard cosmology with cosmological constant, dust and radiation.
Of course, far in the past and future the approximation fails to be accurate.

We can also compute the abundances of dif\/ferent kinds of ef\/fective matter by dividing by the critical density today $\tilde\rho_0^{\rm cr}= 3\tilde H_0=
\rho_0$  to obtain the ef\/fective abundance parameters today
\begin{gather*}
\tilde\Om{}_\La^0 \simeq   \Frac[  \rho_0/8] \ep -\Frac[\rho_0^2/4]\ep^2,
\qquad
\tilde\Om{}^0_d\simeq  1
-\Frac[5 \rho_0/4]\ep
+\Frac[11\rho_0^2/8]\ep^2,
 \qquad
 \tilde\Om{}^0_r\simeq
 \Frac[9  \rho_0/8]\ep
 -\Frac[9\rho_0^2/8]\ep^2.
\end{gather*}

In principle, one can try and determine $\ep$ so that, e.g., $\Om_\La\simeq 0.75$ as observed, e.g., by WMAP.
However, there are numerical evidences that the {(exact)} abundance of ef\/fective dark energy in $f(\calR)=\calR - \sFrac[\ep/2]\calR^2$
models cannot exceed a value of about~0.62.
In this case the whole family of models  $f(\calR)=\calR - \sFrac[\ep/2]\calR^2$ can be rejected since it is not compatible with observations.

\subsection{The model surface}

What we got  above as a description of the cosmological model is the abundance parameters
$\big(\Om^0_k, \Om^0_\La, \Om^0_d, \Om^0_r, \dots\big)$ as functions of the parameter $\ep$ which in fact parametrises
the family of models under consideration, namely {in the simple example of} the function $f(\calR)=\calR - \sFrac[\ep/2]\calR^2$.

Abundances are particularly interesting parameters since they can be measured quite directly observing the Cosmic Microwave Background radiation.
Accordingly, one has  independent ef\/fective abundances  $\Om^0_i$ ($i=1, \dots, n$) which result from observations (the curvature~$\Om^0_k$
is eliminated by the constraint $\sum \Om_i^0=1$).
More generally, a family of models is represented by a number of parameters~$\ep^\al$ (the constants determining the function $f(\calR)$) and the independent abundances
\begin{gather}
\Om^0_i= \Om^0_i(\ep^\al)
\label{Surface}
\end{gather}
def\/ine a map from the set of models into the space of cosmic pies.

By letting the parameters $\ep^\al$ vary,  the equations (\ref{Surface}) describe a parametrised subset $\Si\subset \R^n$. Each point on~$\Si$ represents the predictions of a model in the family.
In this setting one can translate physical issues in geometrical claims, e.g., whether an observed cosmic pie $(\Om^0_i)\in \R^n$ belongs to the image $\Si\subset \R^n$.

The subset $\Si$ is called the {\it model surface} and it captures all the predictions of the models in terms of observable quantities.
Thus if the observed cosmic pie is not on~$\Si$ then the family cannot account for observations and must be abandoned.

\begin{figure}[t]
   \centering
   \includegraphics[width=5cm]{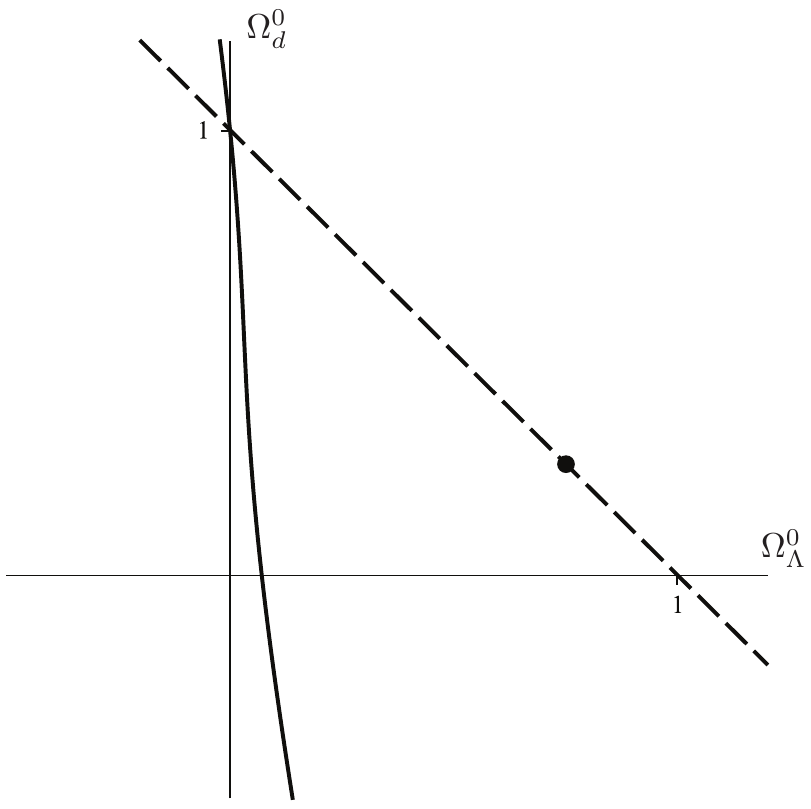}
   \caption{Line model for  $f(\calR)=\calR - \sFrac[\ep/2]\calR^2$. Dash line is the constraint $\sum_i\Om^0_i=1$. The dot is the observered cosmic pie (e.g., WMAP).}
   \label{fig2}
\end{figure}

In the model above, one has one parameter $\ep$ and two independent abundances $\big(\Om^0_\La, \Om^0_d\big)\in \R^2$.
The corresponding line representing the model is shown in Fig.~\ref{fig2}.
Since observations are considered to support  $\big(\Om^0_\La, \Om^0_d\big)\simeq (0.75, 0.25)$
the family is discarded.

More generally, one can consider further observable quantities as the cosmographic parameters
\begin{gather*}
q= -\Frac[1/H^2]\Frac[\ddot a/a],
\qquad
j= \Frac[1/H^3]\Frac[\dot{\ddot a}/a],
\qquad
j= \Frac[1/H^4]\Frac[\ddot{\ddot a}/a],
\qquad \dots
\end{gather*}
in the target space.
On the other hand, ideally, one can expand the function $f(\calR)$ in series (for example in Laurent series) and consider the inf\/initely many coef\/f\/icients as the family parameters $\ep^\al$.
Although, this becomes mathematically elaborated (having an inf\/inite-dimensional surface embedded into an inf\/inite-dimensional space)
one can regard this as a universal setting to represent theories and observations as geometric intrinsic claims even when one may simplify it projecting to f\/inite-dimensional representations in some particular context.
For example, in this context is relatively easy and clear to argue if relations among observables hold in general or are valid within a specif\/ic family of theories when they emerge on the surface.

\section{Conclusions and perspectives}\label{section5}

 In the literature,  Palatini $f(\calR)$-theories have not been studied much because of a number of shortcomings they are believed to have; see, for example,~\cite{R1a} and~\cite{R1b} where it was noticed that Palatini variation is not generally equivalent to metric variation. Since we totally agree on this point, we did not consider metric variations which are {\it different} theories.
For us Palatini theories are motivated by the EPS formalism and they are, in our opinion, more natural than the metric ones.

In the second place, one can perform a conformal on-shell transformation to show dynamical equivalence with {purely metric} Brans--Dicke theories (with a suitable potential); see \cite{DeFelice, Faraoni}. Since Brans--Dicke models (without a potential) are ruled out by solar system observations (e.g., precession of Mercury's perihelia) {it has been claimed that}
then any Palatini $f(\calR)$-theory should be ruled out as well.
This is false due to two issues. First, one should consider the role of the potential.
Second, and more importantly, although f\/ield equations of the two theories are equivalent that does not imply a complete physical equivalence.
In particular, in the case of precession of perihelia, the test on Brans--Dicke models is performed by assuming that planets go along timelike geodesics of~$g$ while in Palatini $f(\calR)$-theories they go along geodesics of~$\tilde g$.
One can show by a detailed analysis that there is a huge family of Palatini $f(\calR)$-theories which in fact pass the test of perihelia; see~\cite{Kepler}.

Moreover, the conformal transformation needed to recast Palatini-$f(\calR)$ theories into Brans--Dicke form is quite peculiar.
One starts with a Palatini theory in which conformal transformations {act} on the metric (which do not touch the connection which is independent) {and they are symmetries}. Then one solves the f\/ield equation related to the connection learning that the connection is in fact the Levi-Civita connection of a~suitably def\/ined conformal metric~$\tilde g$.
One is left with the f\/ield equation depending on the metric (and the conformal factor) alone
since the connection has been eliminated. At this point by performing the inverse conformal transformation on the metric
(which now af\/fects the connection as well) the theory is recast into a~Brans--Dicke form.
There should be no surprise that the two theories are inequivalent since the latter transformation is not a symmetry.

Another, issue is about star models. It was argued that assuming polytropic equation of state (EoS)
one cannot match the inner solution with the outer solution for $g$.
This is in fact true due to the fact that the conformal factor is continuous but not~$\calC^1$ on the surface of the star.
If one matches the conformal metric~$\tilde g$ (how is natural to do in view of the interpretation above) this is possible and the discontinuity appears only in the original metric~$g$.
Moreover, it has been noted that here the issue is due to the assumption about the EoS which is not a fundamental principle.
Allowing a slight deformation of EoS one can in fact match the metrics at the surface of the star.
The same result can be obtained by changing the external solution from a vacuum solution to an almost-vacuum one
as it is reasonable and physically sound; see \cite{noGo1,noGo3, noGo2}.

We showed how to classify extended cosmological models with respect to their predictions about the cosmic pie observations and to their predictions of the evolution of the scale factor.

We showed that, the species revealed in the cosmic pie are essentially a~{\it choice} in cosmology, or, in other words, an input from other physics (structure formation, baryogenesis, \dots).
However, once this input is given, extended gravity models produce  quantitative predictions which allow in principle to falsify or corroborate models.

The choice of ef\/fective species is not forced by mathematics essentially because of the appro\-xi\-ma\-tion method used. We chose to expand~$\tilde \rho$ in negative powers of~$\tilde a$ around the value $\tilde a=1$.
Even if we had $\tilde \rho= 3 \tilde a^{-2}$ and decide we want to see only dust and cosmological factor we would f\/ind
\begin{gather*}
\tilde\rho\sim 1+\Frac[2/{\tilde a}{}^3],
\end{gather*}
which approximates the actual ef\/fective density around today $\tilde a=1$.

One clearly sees that if one has a direct observation of~$a(t)$ in the past and assumes the approximated value
then one can predict at what time in the past the deviation is f\/irst observed.

Let us stress that, although it may not be computationally simple, the predictions of the models for the cosmic pie are extracted {\it before} solving the model.

\begin{figure}[t]
   \centering
  \includegraphics[width=6cm]{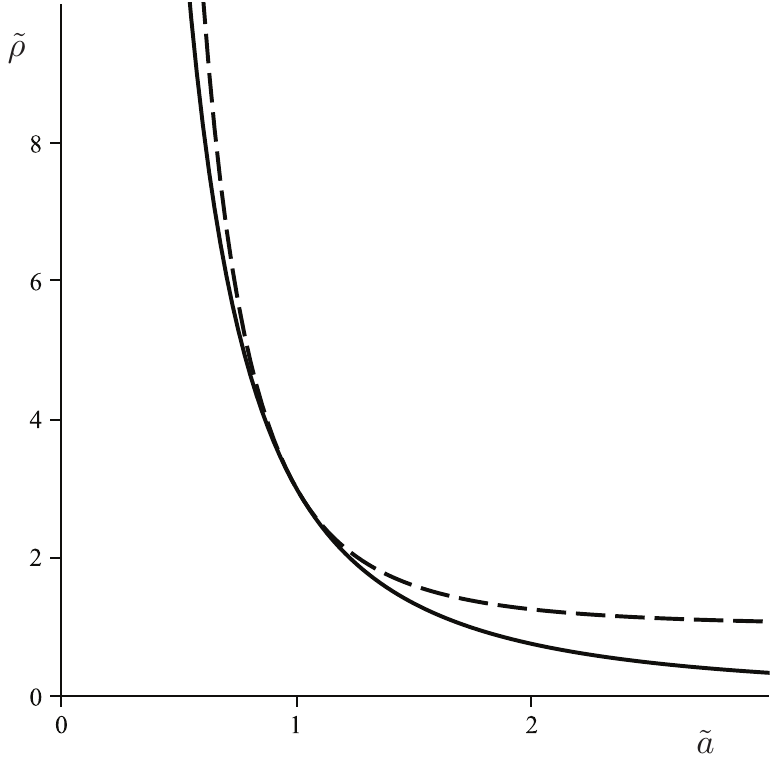}
   \caption{Approximation of a curvature density by cosmological constant and dust (dash line is the approximated density).}
   \label{fig3}
\end{figure}

Future investigations are to be devoted to models in which the conformal factor stays positive everywhere, so that signature changes
are prevented.

{Another research perspective will be investigating how one can encode physical properties (e.g., late time acceleration) as properties of the model surface and viceversa reading physical properties of the theories from the geometry of the surface.
Currently, the main dif\/f\/iculty to be overcome is to control with enough generality the inversion of the master equation which plays a prominent role in expressing the embedding map~(\ref{Embed}).
On the other hand, the model surface may be a good tool exactly because it shows that the properties are {\it algebraically} encoded in the model precisely through the master equation of the theory.}

\appendix

\section{Conservation laws}\label{appendixA}

In a $f(\calR)$-theory, the ef\/fective stress tensor $\tilde T_{\mu\nu}$ is def\/ined by equation~(\ref{EffectiveT}).
Its conservation with respect to the conformal metric~$\tilde g$ reads as $\tilde\na_\mu \tilde T^{\mu\nu}=0$,
where indices have been raised by~$\tilde g$.
The relations between controvariant stress tensors is then
\begin{gather*}
\vp^3 \tilde T^{\mu\nu}= T^{\mu\nu}  -  (\vp \calR -f(\calR) ) g^{\mu\nu}.
\end{gather*}
Let us stress that indices of $\tilde T^{\mu\nu}$ are here raised by~$\tilde g$, while indices of $T^{\mu\nu} $ are raised by~$g$.

Let us introduce the tensor
\begin{gather*}
K^\al_{\be\mu}= \{\tilde g\}^\al_{\be\mu}- \{ g\}^\al_{\be\mu}= -\sFrac[1/2]\big( g^{\al\ep}g_{\be\mu} -2 \de^\al_{(\be}\de^\ep_{\mu)}\big)\nab{\ast}_\ep\ln \vp,
\end{gather*}
which accounts for the dif\/ference between the connections of the metrics~$\tilde g$ and~$g$.
Hereafter the covariant derivatives with respect to~$g$ will be denoted by $\na_\ep$ while the ones with respect to~$\tilde g$
will be denoted by $\tilde \na_\ep$.
The symbol $\nab{\ast}_\ep$ means a covariant derivative when it is independent of any connection (in this case being applied to the scalar~$\ln\vp$).

One can easily show that
\begin{gather*}
K^\mu_{\ep\mu}\tilde T^{\ep\nu} = 2 \tilde T^{\ep\nu}\nab{\ast}_\ep \ln\vp
\end{gather*}
and
\begin{gather*}
K^\nu_{\ep\mu }\tilde T^{\ep\mu} =  \tilde T^{\nu\ep}\nab{\ast}_\ep \ln \vp-\sFrac[1/2] g^{\nu\al} g_{\ep\mu} \tilde T^{\ep\mu}\nab{\ast}_\ep \ln \vp.
\end{gather*}
Then one has directly that
\begin{gather*}
\vp^3 \tilde\na_\mu \tilde T^{\mu\nu}=  \vp^3 \big(\na_\mu \tilde T^{\mu\nu} + K^\mu_{\ep\mu}\tilde T^{\ep\nu} +K^\nu_{\ep\mu }\tilde T^{\ep\mu} \big)  \\
\hphantom{\vp^3 \tilde\na_\mu \tilde T^{\mu\nu}}{}
=  \na_\mu  T^{\mu\nu} -\sFrac[1/2] g^{\nu\al}\nab{\ast}_\al \ln \vp\big( T^{\mu\ep} -f(\calR) g^{\mu\ep}\! - \sFrac[1/2] Tg^{\mu\ep}\big) g_{\mu\ep}
 -g^{\nu\al}\nab{\ast}_\al f(\calR) -\sFrac[1/2] g^{\nu\al}\nab{\ast}_\al T \\
\hphantom{\vp^3 \tilde\na_\mu \tilde T^{\mu\nu}}{}
=  \na_\mu  T^{\mu\nu} + g^{\nu\al} \big( f''(\calR )\calR \nab{\ast}_{\al} \calR
-f'(\calR )\nab{\ast}_{\al} \calR -\sFrac[1/2] \nab{\ast}_{\al} T\big).
\end{gather*}

However, by taking the covariant derivative of the master equation (\ref{MasterEquation}) one has
\begin{gather*}
f''(\calR )\calR \nab{\ast}_{\al} \calR
+f'(\calR )\nab{\ast}_{\al} \calR-2f'(\calR )\nab{\ast}_{\al} \calR =\sFrac[1/2] \nab{\ast}_{\al} T,
\end{gather*}
so that one simply has
\begin{gather*}
\vp^3 \tilde\na_\mu \tilde T^{\mu\nu}=\na_\mu  T^{\mu\nu},
\end{gather*}
so that $ \tilde T^{\mu\nu}$ is conserved with respect to $\tilde g$ if\/f $T^{\mu\nu}$ is conserved with respect to $g$.

\section{Model solution}\label{appendixB}

Solving the FRW equations may be dif\/f\/icult though in a sense trivial since, in the end,
one has to solve the Weierstrass equations with
\begin{gather*}
\tilde \Phi (\tilde a)= -k + \Frac[1/3] \tilde\rho(a(\tilde a)) \tilde a^2,
\end{gather*}
i.e., computing the integral
\begin{gather*}
\tilde t=\tilde t_0+\int_{\tilde a_0}^{\tilde a} \Frac[d\tilde a/\sqrt{\tilde \Phi (\tilde a)}].
\end{gather*}
Maple can make it (with some ef\/fort) numerically (which means one can obtain a plot of \mbox{$\tilde a = \tilde a(\tilde t)$)}.

An easier way is to change variable back to  $(t, a)$, solve the equation and then compute the conformal factor, the new time and the conformal scale factor.
This can be done by considering that
\begin{gather*}
\Frac[d\tilde a/d\tilde t]=\Frac[d\tilde a/d a] \Frac[d a/d t] \Frac[d t/d\tilde t]
\qquad\then\qquad
\left(\Frac[d a/d t] \right)^2 = \left( { \Frac[d \tilde t/d t] }\right)^2 \left({\Frac[d\tilde a/d a]  }\right)^{-2}\tilde \Phi (\tilde a) =: \Phi(a).
\end{gather*}
Since we know that
$d\tilde t=\sqrt{\vp} dt $ and $\tilde a(a)= a \sqrt{\vp}$  one has for $f(\calR)=\calR-\sFrac[\ep/2]\calR^2$
\begin{gather*}
\Phi(a) :=  \left( \Frac[3-\ep \rho(a)/3-\ep \rho_0] \right) \left({\Frac[d\tilde a/d a] (a) }\right)^{-2} \tilde \Phi (\tilde \rho(a)).
\end{gather*}

Once we have $\Phi(a)$ we obtain the time coordinate as
\begin{gather*}
t(a)=t_0 + \int_{a_0}^{a} \Frac[da / \sqrt{\Phi(a)}]
\end{gather*}
and the conformal time as
\begin{gather*}
\tilde t(a)=\tilde t_0 + \int_{t_0}^{t} \sqrt{\vp(a(t))} dt = \tilde t_0 + \int_{a_0}^{a} \sqrt{\vp(a)} \Frac[da / \sqrt{\Phi(a)}].
\end{gather*}
Finally the original density $\rho$ is given as $\rho(a)$ by EoS as well as the conformal density and pressure $(\tilde \rho(a), \tilde p(a))$.

\begin{figure}[t]
   \centering
  \includegraphics[width=5cm]{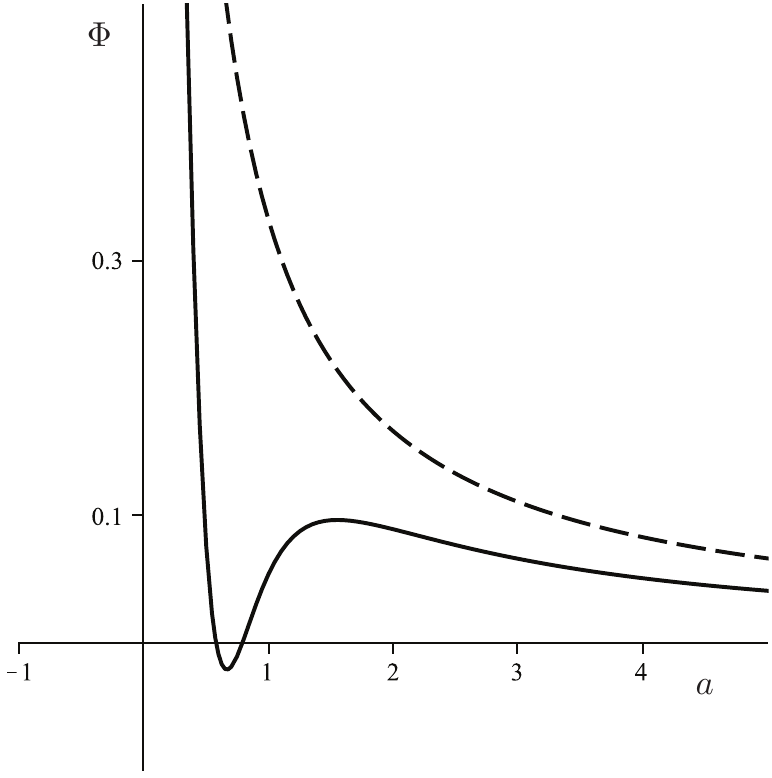}
   \caption{Weierstrass function $\Phi(a)$. (Dash line is standard GR.)}
   \label{fig4}
\end{figure}

 In our special case  $f(\calR)=\calR-\sFrac[\ep/2]\calR^2$, setting $\ep=\sFrac[1/2]$ and $\rho_0=1$ , one obtains (see Fig.~\ref{fig4})
\begin{gather*}
\Phi(a)=  \Frac[1/3a] \Frac[(2a^3-1)(5a^3-1)/(4a^3+1)^2]
\end{gather*}
with two allowed regions, one being $a\in [0, 5^{-1/3}]$ the other being $a\ge 2^{-1/3}$.
The f\/irst region corresponds to an oscillating behaviour, hence to a~closed universe.
The second region corresponds to an expanding universe.
Let us consider the second region.

\begin{figure}[t]
   \centering
  \includegraphics[width=5cm]{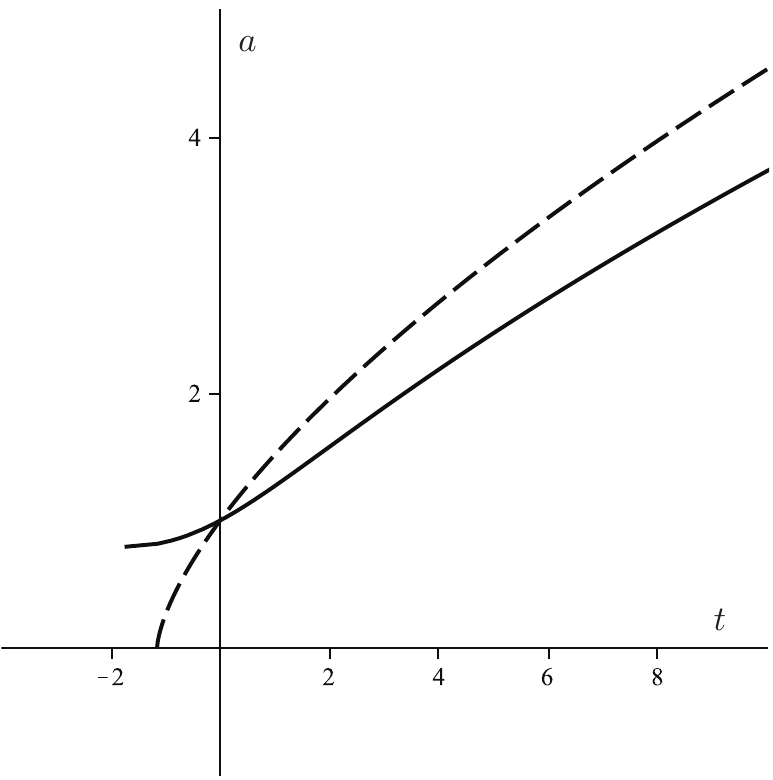}
  \qquad
  \includegraphics[width=5cm]{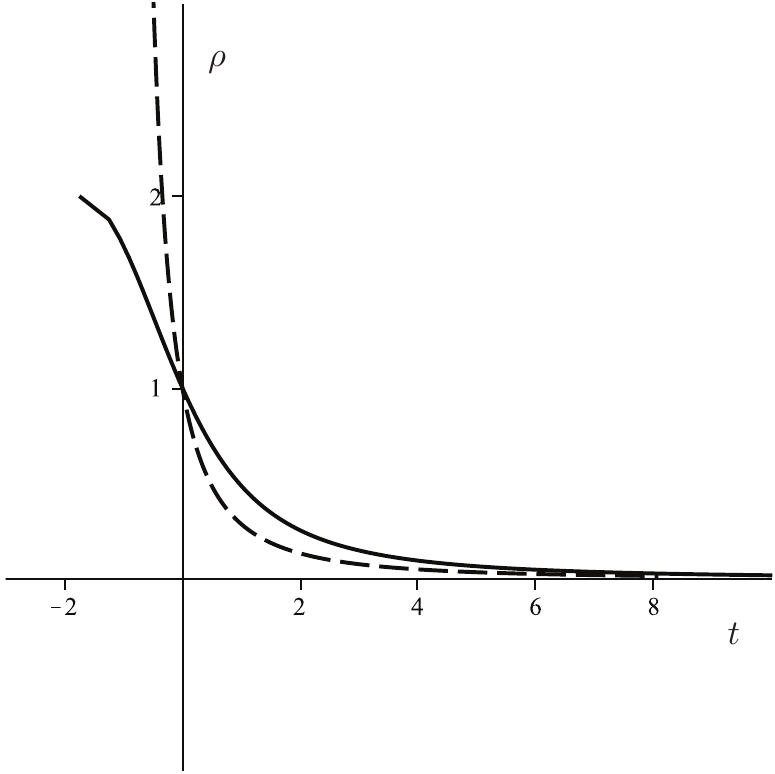}
   \caption{a)~Scale factor~$a$  as a function of~$t$ (dash line standard~GR).
b)~Energy density~$\rho$ for visible matter as a function of~$t$ (dash line standard GR).}
   \label{fig5}
\end{figure}

The original scale factor hence appears as a quite natural parameter to (more) easily integrate equations for~$\tilde a$.

In this way one is able to plot parametrically everything; see Fig.~\ref{fig5}.
Notice how the energy density of the visible matter $\rho(t)$ stays bounded and the scale factor~$a(t)$ stays non-zero when standard GR has a singularity.

This means that the solution can be extended to negative~$t$ and the big bang singularity is avoided (as some model of quantum gravity predicts; see~\cite{LQC}).

\subsection*{Acknowledgements}

We are grateful to A.~Borowiec for comments and discussion.
We acknowledge the contribution of INFN (Iniziativa Specif\/ica QGSKY),
the local research project {\it Metodi Geometrici in Fisica Matematica e Applicazioni} (2015) of Dipartimento di Matematica of University of Torino (Italy).
This paper is also supported by INdAM-GNFM.

\pdfbookmark[1]{References}{ref}
\LastPageEnding

\end{document}